\documentclass[twocolumn,shortnote]{jpsj3}
\usepackage{bm}

\def\gsim{\mathop {\vtop {\ialign {##\crcr
$\hfil \displaystyle {>}\hfil $\crcr \noalign {\kern1pt \nointerlineskip
 }
$\,\sim$ \crcr \noalign {\kern1pt}}}}\limits}
\def\lsim{\mathop {\vtop {\ialign {##\crcr
$\hfil \displaystyle {<}\hfil $\crcr \noalign {\kern1pt \nointerlineskip
 }
$\,\,\sim$ \crcr \noalign {\kern1pt}}}}\limits}

\title{ New Magnetic Phases in the Chiral Magnet CsCuCl$_3$ under High Pressures}

\author{ Masashi {\sc Hosoi}\thanks{hosoi@hosi.phys.s.u-tokyo.ac.jp}, Hiroyasu {\sc Matsuura}, and Masao {\sc Ogata}
}
\inst
{
Department of Physics, University of Tokyo, 7-3-1 Hongo, Bunkyo, Tokyo 113-0033, Japan\\
}

\recdate
{
\today
}

\abst
{ We study the magnetic phase diagram of CsCuCl$_3$ within spin-wave theory.
 We clarify the existence of new magnetic phases, namely the \textit{up-up-down} and Y coplanar phases under a high pressure.
We also discuss a magnetic field$(H)$- temperature$(T)$ phase diagram under ambient and high pressures.  }

\begin{document}
\sloppy

\maketitle
ABX$_3$-type compounds (A=Rb or Cs; B=Mn, Fe, Co, Ni, Cu, or V; X=Cl, Br, or I) provide an ideal platform for investigating low-dimensional phase transitions.
Within that family, CsCuCl$_3$ has attracted much attention with regard to two-dimensional triangular antiferromagnets and chirality.
The crystal structure of CsCuCl$_3$ at low temperatures (below 423 K) is hexagonal with the space group $P6_122$ or $P6_522$. In this phase, Cu chains
form helices along the $c$ axis with six Cu atoms per unit cell.
These chains form a triangular lattice in the $ab$ plane.~\cite{Adachi}
The main exchange interactions between the Cu ions are the intra-chain ferromagnetic interactions (coupling constant $J_0\sim28$ K), inter-chain antiferromagnetic interaction (coupling constant $J_1\sim4.9$ K), and Dzyaloshinskii--Moriya (DM)
interaction with the $\mbox{\boldmath $D$}$ vector pointing along the $c$ axis ($|\bm{D}|\sim5$ K).~\cite{Adachi,Kousaka} Because of these interactions, below $T_N = 10.7$K ,
 this compound displays a helical 120$^\circ$ spin structure along the $c$ axis with a pitch angle of $\theta = 5.1^\circ$.~\cite{Adachi}

The ground state of CsCuCl$_3$ in a longitudinal magnetic field ($\mbox{\boldmath $H$}\parallel c$) at a low temperature and under ambient pressure is well understood within spin-wave theory.
Nikuni and Shiba showed that the quantum-phase transition from an umbrella phase to a 2-1 coplanar phase occurs when the magnetic field increases.~\cite{Nikuni}
This theoretical prediction was confirmed by neutron diffraction and specific heat measurements.~\cite{Mino,Weber}
On the other hand, a new magnetization plateau was recently found under high pressure.~\cite{Sera}
Although this plateau is expected to be an \textit{up-up-down} (uud) phase showing the 1/3 plateau of the saturation magnetic field, its existence has so far not been explained within spin-wave theory.~\cite{Nikuni}

In this paper, we predict the existence of the uud phase theoretically by considering the pressure dependence of exchange interactions, on the basis of spin-wave theory.
We also predict a Y coplanar phase under high pressure, which has not been confirmed experimentally.
We then also examine thermal fluctuations for each phase, and the $H$-$T$ phase diagram.

We write the Hamiltonian of CsCuCl$_3$ in $\mbox{\boldmath $H$}\parallel c$ as
\begin{align*}
  {\mathcal H} &= -2J_0\sum_{i,n}(\mbox{\boldmath $S$}_{i,n}\cdot\mbox{\boldmath $S$}_{i,n+1}+\eta(S_{i,n}^xS_{i,n+1}^x+S_{i,n}^yS_{i,n+1}^y))\\
  &+2J_1\sum_{\langle ij\rangle,n}\mbox{\boldmath $S$}_{i,n}\cdot\mbox{\boldmath $S$}_{j,n}-\sum_{i,n}\mbox{\boldmath $D$}_{n,n+1}\cdot
  (\mbox{\boldmath $S$}_{i,n}\times\mbox{\boldmath $S$}_{i,n+1})\\
  &-g\mu_BH\sum_{i,n}S_{i,n}^z. \tag{1}
\end{align*}
Here, $\mbox{\boldmath $S$}_{i,n} \equiv (S_{i,n}^x, S_{i,n}^y, S_{i,n}^z)$ is a spin operator at the $i$-th site in the $n$-th $ab$ plane, and the summation $\langle ij \rangle$ covers nearest-neighbor sites in the
$ab$ plane. The $z$ axis is parallel to the magnetic field $H$ and $\mbox{\boldmath $D$}_{n,n+1} \equiv (0,0,D)$ is the DM interaction between the $i$-th spins in the $n$-th and $(n+1)$-st planes.
The quantity $\eta$ is an anisotropic exchange interaction of the easy-plane type, and
$g$ and $\mu_B$ are the $g$-factor ($g=2$) and Bohr magneton, respectively.

We can eliminate the DM interaction term by rotating the $xy$ plane about $z$ axis by an angle $q = \tan^{-1}(D/2J_0(1+\eta))$.~\cite{Nikuni} Then, the Hamiltonian is rewritten
\begin{align*}
  {\mathcal H}' &= -\sum_{i,n}(2\tilde{J_0}(S_{i,n}^xS_{i,n+1}^x+S_{i,n}^yS_{i,n+1}^y)+2J_0S_{i,n}^zS_{i,n+1}^z)\\
  &+2J_1\sum_{\langle ij\rangle,n}\mbox{\boldmath $S$}_{i,n}\cdot\mbox{\boldmath $S$}_{j,n}-g\mu_BH\sum_{i,n}S_{i,n}^z, \tag{2}
\end{align*}
where $\tilde{J_0} = J_0\sqrt{(1+\eta)^2+(D/2J_0)^2}$. Note that the DM interaction is incorporated into the easy-plane anisotropy and we define the anisotropy parameter as
\begin{align*}
\Delta \equiv (\tilde{J_0}-J_0)/J_1. \tag{3}
\end{align*}

We consider the five spin configurations shown in Fig. 1 as candidates for the ground state.
\begin{figure}[h]
\begin{center}
\rotatebox{0}{\includegraphics[width=\linewidth]{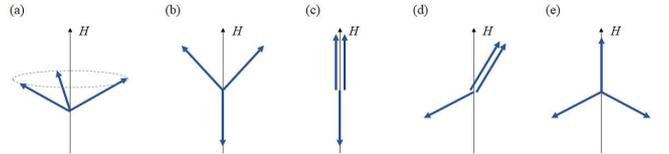}}
\caption{(Color online) Five spin configurations considered as candidates for the ground state: (a) umbrella, (b) Y coplanar, (c) uud,
(d) 2-1 coplanar, and (e) inverse Y configurations.}
\label{Fig1}
\end{center}
\end{figure}
The umbrella configuration (a) is selected for finite $\Delta$ in the classical spin model and the saturation field $H_s$ is obtained as~\cite{Nikuni}
\begin{align*}
 H_s=(18+4\Delta)J_1S/g\mu_B.  \tag{4}
\end{align*}
Then, we consider the quantum fluctuations using spin-wave theory. By performing a calculation similar to that in ref. [\citen{Nikuni}], we obtain the bosonic Bogoliubov-de Gennes Hamiltonian
\begin{align*}
  {\mathcal{H}}_{\mathrm{SW}}=&E_0-(2J_0+3J_1)NS\\
  &+\frac{S}{2}\sum_{\bf k} (\mbox{\boldmath $a$}_{\bf k}^\dagger\ \mbox{\boldmath $a$}_{-{\bf k}})
  \begin{pmatrix}
    E_{\bf k} & F_{\bf k} \\
    F_{-{\bf k}}^* & E_{-{\bf k}}^*
  \end{pmatrix}
  \begin{pmatrix}
    \mbox{\boldmath $a$}_{\bf k} \\
    \mbox{\boldmath $a$}_{-{\bf k}}^\dagger
  \end{pmatrix}
  .
  \tag{5}
\end{align*}
Here, $E_0$ is the classical energy, $N$ is the total number of spins, $\mbox{\boldmath $a$}_{\bf k}=(a_{{\bf k},1}, a_{{\bf k},2}, a_{{\bf k},3})$ is a three-component vector of annihilation
operators corresponding to three sublattices, and $E_{\bf k}$ and $F_{\bf k}$ are $3\times3$ Hermitian matrices.
The components of these matrices are denoted
\begin{align*}
  E_{jj} &= 4\tilde{J_0}(1-\cos k_z)+6J_1-2J_1\Delta\sin^2\theta_j\cos k_z,  \tag{6}\\
  F_{jj} &= 2J_1\Delta\sin^2\theta_j\cos k_z,   \tag{7}
\end{align*}
and the other matrices are the same as those in ref. [\citen{Nikuni}].
Here, $\theta_j$ is the $j$-th polar angle from the classical spin direction.
In ref. [\citen{Nikuni}], the terms including $\Delta$ of eqs. (6) and (7) were neglected because they are small.
However, we presently consider these terms to study the pressure dependence of $\Delta$, as discussed below.

We can diagonalize ${\mathcal{H}}_{\mathrm{SW}}$ by using the bosonic Bogoliubov transformation~\cite{Shindou}. The total energy is obtained as
\begin{equation*}
 E_{\mathrm{SW}}=E_0 -(2J_0+3J_1)NS+\frac{S}{2}\sum_{ i,{\bf k}}\omega_i({\bf k}),\tag{8}
\end{equation*}
where $\omega_i({\bf k})$ is the $i$-th eigenvalue with wave vector ${\bf k}$.
We investigated the spin configuration in the ground state by comparing the total energy $E_{\mathrm{SW}}$ for each spin configuration.
As the pressure increases, we consider the decrease in the ratio of the lattice constants in the $a$ and $c$ directions.~\cite{Christy}
The anisotropy parameter $\Delta$ decreases, which we attribute to the increasing pressure.

Figure 2 shows the $h$-$\Delta$ phase diagram obtained for CsCuCl$_3$  at $T = 0$ in the longitudinal magnetic field ($h=H/H_s$).
The parameters are set to $J_0 = 28$ K, $\tilde{J_0}=1.012J_0$, and the amplitude of $\Delta$ is related to the applied pressure.
Under ambient pressure, $\Delta$ is estimated to be approximately 0.07~\cite{Tanaka}, as shown in ref. [\citen{Nikuni}], and a phase transition between the umbrella and 2-1 coplanar phases is observed.
On the other hand, in the high-pressure region (small $\Delta$), the Y coplanar phase and uud phase shown in Fig. 1(b) and (c) appear between the umbrella and 2-1 coplanar phases.
This result is consistent with experiments~\cite{Sera} if we consider that the experiment performed under $P=0.68$ GPa corresponds to $\Delta \simeq 0.06$.
As for the Y coplanar phase, the signal of phase transition is unclear in the magnetization measurement of ref [\citen{Sera}].
High-pressure measurements are therefore needed to confirm this theoretical result.
\begin{figure}[h]
\begin{center}
\rotatebox{0}{\includegraphics[width=0.8\linewidth]{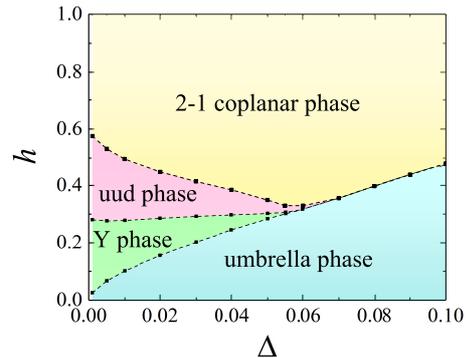}}
\caption{(Color online) The $h$-$\Delta$ phase diagram of CsCuCl$_3$ in a longitudinal magnetic field at $T=0$. The anisotropy $\Delta$
corresponds to the pressure and $h$ corresponds to the magnetic field strength.}
\label{Graph1}
\end{center}
\end{figure}

To investigate the properties at finite temperatures, we compare the free energy of each spin configuration
\begin{align*}
  F = E_{\mathrm{tot}}+T\sum_{i=1}^3\sum_{\bf k}\ln(1-e^{-S\omega_i({\bf k})/T}). \tag{9}
\end{align*}
We show the obtained $h$-$T$ phase diagram under ambient pressure ($\Delta = 0.07$) in Fig. 3 (a).
Although the uud phase is not observed at $T=0$, it appears at high temperatures.
This is because we neglect the magnon-magnon interaction, which provides an important handle for determining the spin configuration at finite temperatures.
\begin{figure}[h]
\begin{center}
\rotatebox{0}{\includegraphics[width=\linewidth]{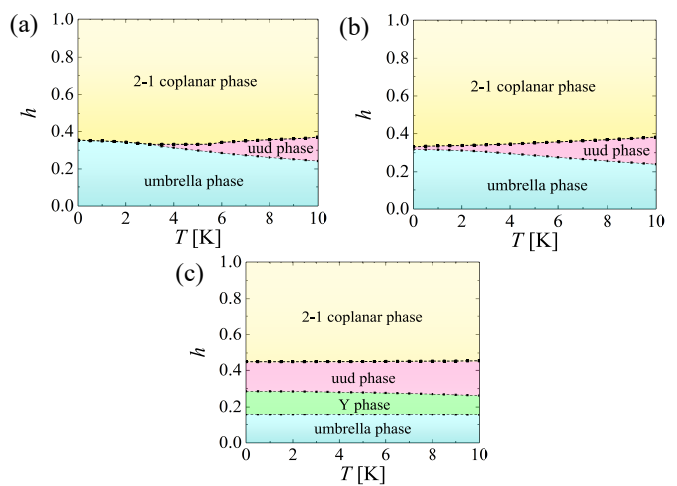}}
\caption{(Color online) The $h$-$T$ phase diagram of CsCuCl$_3$ in a longitudinal magnetic field for (a) $\Delta = 0.07$, (b) $0.06$ and (c) $0.02$ .}
\label{Graph2}
\end{center}
\end{figure}
The $h$-$T$ phase diagram for the high-pressure region ($\Delta = 0.06$) is similar to that under ambient pressure, as shown in Fig.3 (b). In contrast,
the phase diagram for $\Delta = 0.02$ (Fig. 3 (c)) is completely different from those for $\Delta = 0.07$ and $0.06$.
The transition magnetic field is entirely almost temperature-independent up to 10 K, as shown in Fig. 3 (c).

In conclusion, we studied the magnetic-phase diagram of CsCuCl$_3$ in a longitudinal magnetic field under ambient and high pressure within spin-wave theory.
We predicted uud and Y coplanar phases that were not obtained in the previous study.
Furthermore, we showed that the uud phase can appear even under ambient pressure at finite temperature, although we neglect the magnon-magnon interaction.

We acknowledge many fruitful discussions with A. Sera and Y. Kousaka.
This work was supported by the JSPS Core-to-Core Program, A. Advanced Research Net-works. We were also supported by Grants-in-Aid for Scientific Research from the Japan Society for the
Promotion of Science (Nos. 15K17694, 25220803, 17H02912, 17H02923, 18K03482, and 18H01162). M.H. was supported by the Japan Society for the Promotion of Science through the Program for Leading Graduate Schools (MERIT).

\end{document}